\documentclass[12pt]{iopart}
\usepackage{iopams}
\usepackage{graphicx}
\begin{document}
\title{Effect of spin-orbit coupling on tunnelling escape of Bose-Einstein
condensate}

\author{JieLi Qin}
\address{School of Physics and Electronic Engineering, Guangzhou University,
230 Wai Huan Xi Road, Guangzhou Higher Education Mega Center, Guangzhou 510006, China}
\eads{\mailto{qinjielil@126.com}, \mailto{104531@gzhu.edu.cn}}

\vspace{10pt}
\begin{indented}
\item[] December 2018
\end{indented}

\begin{abstract}
We theoretically investigate quantum tunnelling escape of a spin-orbit (SO) coupled
Bose-Einstein condensate (BEC) from a trapping well. The condensate
is initially prepared in a quasi-one-dimensional harmonic trap. Depending
on the system parameters, the ground state can fall in different phases
--- single minimum, separated or stripe. Then, suddenly the trapping
well is opened at one side. The subsequent dynamics of the condensate
is studied by solving nonlinear Schr\"{o}dinger equations. We found
that the diverse phases will greatly change the tunneling escape behavior
of SO coupled BECs. In single minimum and separated phases, the condensate
escapes the trapping well continuously, while in stripe phase it escapes
the well as an array of pulses. We also found that SO coupling has
a suppressing effect on the tunnelling escape of atoms. Especially,
for BECs without inter-atom interaction, the tunnelling escape can
be almost completely eliminated when the system is tuned near the transition
point between single minimum and  stripe phase. Our work thus suggests
that SO coupling may be a useful tool to control the tunnelling dynamic
of BECs, and potentially be applied in realization of atom lasers
and matter wave switches.
\end{abstract}

%
\vspace{2pc}

\submitto{\JPB}

%
%

\section{Introduction}
Quantum tunnelling, the phenomenon where a particle penetrates and
in most cases passes through a potential barrier which it classically
cannot surmount, is one of the most surprising effects in quantum
mechanics. In the early days, it is mainly studied in microscopic level
\cite{Merzbacher2002,Razavy2003}.
In pace with the discovery and development of various macroscopic quantum systems,
macroscopic quantum tunnelling (for example, Josephson tunnelling \cite{Josephson1974})
also begins to attract great research interest \cite{Razavy2003,Takagi2002}.
Specially, Bose-Einstein condensate (BEC) which is macroscopic matter
wave with very good coherence and manipulability is very suitable
for studying quantum tunnelling (and many other macroscopic quantum
phenomena). Atomic Josephson junctions have been realized a long time
ago \cite{Cataliotti2001,Albiez2005}. Furthermore, atomic analogy
of superconducting quantum interference devices (SQUID) has also been
experimentally demonstrated and proposed to be a good compact rotation
sensor \cite{Ryu2013}. Tunnelling escape of BECs from trapping well
has also been extensively studied \cite{Salasnich2001,Carr2005,Fleurov2005,
Huhtamaki2007,Dekel2009,Dekel2010,Alcala2018}
and recently has been experimentally observed as well \cite{Potnis2017,Zhao2017}.

Interference plays a key role in quantum physics. It can have significant
influence on quantum tunnelling. These influences
have been reported in various fields. Fundamentally, it can lead to
breakdown of the exponential decay law \cite{Razavy2003,Winter1961,Gopych1988,Peres1980},
and this phenomenon has already been observed in BEC systems \cite{Wilkinson1997}.
It can also give rise to a modification of tunneling time \cite{Martinez2007}.
In nanomechanical system, it can cause a suppression of the tunneling
between opposite magnetizations \cite{Kovalev2011}. In atomic and
molecular physics, recently it is reported to have significant effect
on the process of photoassociation \cite{Blasing2018} (which is also a quantum
tunnelling related phenomenon).

Spin-orbital (SO) coupling will greatly enrich the quantum properties
of BECs. There exist diverse ground state phases in which the matter
wave properties are quite different \cite{Ho2011,Sinha2011,Li2012,Achilleos2013}.
In zero momentum phase, the condensate stays in a state with zero
momentum, while in separated phase it possess momentum of finite value
either $+p_{0}$ or $-p_{0}$. And specially in stripe phase, its
state is an superposition of $+p_{0}$ and $-p_{0}$ states. The simultaneous
existence of these two states will cause interference between them.
Subsequently, the quantum tunneling phenomena of SO coupled BECs show
new features. In recent years, Josephson tunnelling of SO coupled
BEC attracts much research interests \cite{Zhang2012_043609,March2014,Niu2016,Wang2017,Kartashov2018}.
A new kind of Josephson effect --- momentum space Josephson effect has been predicted \cite{Hou2018}.
It is also reported that SO coupling will lead to rich dynamical phenomena
of BEC Josephson vortices \cite{Toikka2017,Gallem2016}.

However, tunnelling escape of a SO coupled BEC from a trapping well
has not been carefully examined yet. In this work, we will study this
phenomenon, and try to find out what effect will SO coupling have
on the system. Tunnelling escape dynamics of SO coupled BEC in different
phases will be studied by solving nonlinear Schr\"{o}dinger equations,
and the results will be compared. The dependence of tunnelling escaped
atoms number on SO coupling strength will examined. Possible applications
of the system in the field of atom optics will also be discussed.

The following contents of this article is organized as follows. In
section \ref{sec:model}, we introduce the model for studying tunnelling
escape of SO coupled BEC.
 Then in section \ref{sec:results}, tunnelling
escape dynamic of the system is studied both analytically and numerically,
and the results are discussed. At last, the work is summarized in
section \ref{sec:summary}.

\section{Model\label{sec:model}}

\begin{figure}
\begin{centering}
\includegraphics{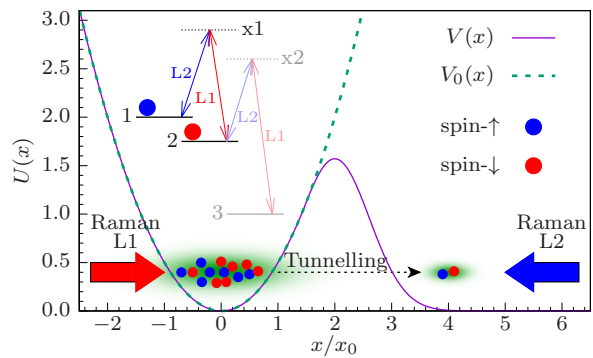}
\par\end{centering}
\caption{Schematic diagram for the tunnelling escape of a SO coupled BEC from
a trapping well. As in reference \cite{Lin2011}, Zeeman levels 1 and 2 act as two spin states
(the third Zeeman level 3 is far detuned and can be adiabatically
eliminated, therefore corresponding lines are plotted in a lighter
color). And SO coupling is realized by interaction between counter-propagating
Raman beams L1, L2 and the atomic condensate. Initially,
the condensate is prepared in the ground state of a harmonic trap
described by equation (\ref{eq:V0}) (green dashed line). At time
$t=0$ the trap is suddenly opened at the right side, i.e., changed
to the form of equation (\ref{eq:V}) (violet solid line) with parameters
$V_{0}=1.57 \hbar\omega_{0}$, $x_{b}=1.5 x_0$, $x_{c}=2 x_0$, $\delta=0.87 x_0$. Then, atoms
begin to escape the trapping well due to quantum tunnelling.
\label{fig:Diagram}}
\end{figure}

We consider a system of SO coupled Bose-Einstein condensate realized
by Raman coupling scheme \cite{Lin2011}, as schematically shown
in figure \ref{fig:Diagram}. The condensate is assumed to be confined
in a parabolic trap with frequencies $\omega_{x}\ll\omega_{\bot}$,
so that the system can be reduced to one-dimension \cite{Salasnich2002},
and the dynamic can be described by the following two coupled nonlinear
Schr\"{o}dinger equations
\begin{eqnarray}
i\hbar\frac{\partial\psi_{\downarrow}}{\partial t} & = & \left[\frac{\left(p_{x}+p_{c}\right)^{2}}{2m}+U\left(x\right)\right]\psi_{\downarrow}+\hbar \Omega\psi_{\uparrow}\nonumber \\
 & + & \left(g_{0}\left|\psi_{\downarrow}\right|^{2}+g_{1}\left|\psi_{\uparrow}\right|^{2}\right)\psi_{\downarrow},\label{eq:GPdown}
\end{eqnarray}
\begin{eqnarray}
i\hbar\frac{\partial\psi_{\uparrow}}{\partial t} & = & \left[\frac{\left(p_{x}-p_{c}\right)^{2}}{2m}+U\left(x\right)\right]\psi_{\uparrow}+\hbar \Omega\psi_{\downarrow}\nonumber \\
 & + & \left(g_{1}\left|\psi_{\downarrow}\right|^{2}+g_{0}\left|\psi_{\uparrow}\right|^{2}\right)\psi_{\uparrow},\label{eq:GPup}
\end{eqnarray}
where $\psi_{\downarrow}$ and $\psi_{\uparrow}$ are wave functions
of the two spin components obeying normalization condition ${\displaystyle \int\left(\left|\psi_{\downarrow}\right|^{2}+\left|\psi_{\uparrow}\right|^{2}\right)dx}=1$,
$\hbar$ is reduced Planck's constant , $m$ is the mass of atom,
$p_{x}=-i\hbar\frac{\partial}{\partial x}$ is the momentum operator
in $x$-direction, $p_{c}$ is the SO coupling strength determined
by momentum transfer of Raman lasers, $\Omega$ is the Rabi coupling
strength accounting for the transition between the two spin states,
$g_{0},g_{1}$ are effective one-dimensional contact interaction strength
(their values are $g_{i}=2N_{0}a_{i}\hbar\omega_{\bot}$, where $a_{0}$
and $a_{1}$ are s-wave scattering lengths for collision of two atoms
with same or different spins, $N_{0}$ is number of atoms in the condensate),
and $U\left(x\right)$ is the external trap potential. Experimentally,
such a system is highly controllable, SO coupling can be tuned by
manipulating the two Raman lasers \cite{Zhang2013,Garcia2015}, and
contact interaction can be tuned using the well-known Feshbach resonance
technique \cite{Chin2010}.

Initially, the condensate is confined in harmonic trap
\begin{equation}
V_{0}\left(x\right)=\frac{1}{2}m\omega_{0}^{2}x^{2}.\label{eq:V0}
\end{equation}
Thus, the initial state is assumed to be ground state $\Psi_{0}$=$\left(\psi_{\downarrow,0},\psi_{\uparrow,0}\right)^{T}$
of this trapping potential. For vanishing of both SO coupling and
inter-atom interactions ($p_{c}=g_{0}=g_{1}=0$), equations (\ref{eq:GPdown})
and (\ref{eq:GPup}) are simplified to
\begin{equation}
\mu\varphi_{\downarrow,\uparrow}=\left(\frac{p_{x}^{2}}{2m}+V_{0}\left(x\right)\right)\varphi_{\downarrow,\uparrow}+\hbar \Omega\varphi_{\uparrow,\downarrow},\label{eq:OnlyRabi}
\end{equation}
with $\psi_{\downarrow,\uparrow}\left(x,t\right)=\varphi_{\downarrow,\uparrow}\left(x\right)e^{-i\mu t/\hbar}$.
Solving this time-independent Schr\"{o}dinger equation, the ground
state can be calculated precisely to be a Gaussian wave packet
\begin{equation}
\Phi_{0}^{G}=\frac{1}{\sqrt{2\sqrt{\pi}x_{0}}}\left(1,-1\right)^{T} \exp\left[-\frac{x^{2}}{2x_{0}^{2}}\right].\label{eq:Gaussian}
\end{equation}
Otherwise, for no vanishing of SO coupling and inter-atom interaction,
the initial state will be obtained by numerically evolving a trial wave
function in imaginary time dimension.

The ground state properties of SO coupled BEC have been carefully studied in
reference \cite{Li2012}. Here we give a brief review. Omitting the
nonlinear and trapping terms, at the same time assuming a plane wave
solution $\psi_{\downarrow,\uparrow}\left(x,t\right)=\psi_{\downarrow,\uparrow}\exp\left[i\left(p_{x}x-Et\right)/\hbar\right]$,
the dispersion relation of SO coupled matter wave can be obtained
by diagonalizing equations (\ref{eq:GPdown}) and (\ref{eq:GPup}),
which reads $E_{\pm}\left(p_{x}\right)=\left(p_{x}^{2}+p_{c}^{2}\right)/\left(2m\right)\pm\sqrt{\hbar^{2}\Omega^{2}+p_{c}^{2}p_{x}^{2}/m^{2}}.$
This dispersion relation splits into two branches --- $E_{+}\left(p_{x}\right)$
and $E_{-}\left(p_{x}\right)$. The higher energy branch $E_{+}\left(p_{x}\right)$
is quite normal. It always has a single minimum at $p_{x}=0$, and
will have little influence on ground state property of the system. Interestingly,
the lower energy branch $E_{-}\left(p_{x}\right)$ behaves very differently
for different strength of SO and Rabi coupling. When $\hbar\Omega>p_{c}^{2}/m$,
it also has only one minimum at $p_{x}=0$. Therefore, the lowest
energy state will carry zero momentum, and the corresponding phase
is called ``zero momentum phase''. While in the case of $\hbar\Omega<p_{c}^{2}/m$,
the lower energy branch dispersion curve has two equal minima at $p_{x}=-p_{0},+p_{0}$
(with $p_{0}=\sqrt{p_{c}^{2}-\left(m\hbar\Omega/p_{c}\right)^{2}}$).
As a result, the lowest energy state may carry separated momentum
$-p_{0}$ or $+p_{0}$ (separated phase), or a superposition of them
(due to the interference between these two states, there exist interference
stripes, thus the corresponding phase is called
``stripe phase''), and these states are degenerate. This degeneracy
may be lifted by the external potential or interaction between atoms.
When there is only the external harmonic trap potential, the system
prefers to stay in stripe phase. And when interaction is considered,
as the SO coupling strength getting strong, the system can firstly
approach separated phase, then reaches the stripe phase.
In the following content, these conclusions will be used directly
without more explanation.

At time $t=0$ trapping potential (\ref{eq:V0}) is suddenly opened
at the right side. Mathematically speaking, the trapping potential
is suddenly changed to the following form (see figure \ref{fig:Diagram})
\begin{equation}
V\left(x\right)=\left\{ \begin{array}{c@{\quad}r}
m\omega_{0}^{2}x^{2}/2,  &x\leq x_{b},\\
V_{0}\exp\left[-\left(\frac{x-x_{c}}{\delta}\right)^{2}\right],  &x>x_{b}.
\end{array}\right.\label{eq:V}
\end{equation}
Here parameters $V_{0},x_{c},\delta$ are determined by matching the
potential function and its first derivative at point $x=x_{b}$. They
represent height, location and width of the barrier respectively,
and their values are set to $V_{0}=1.57 E_0$, $x_{b}=1.5x_{0}$,
$x_{c}=2x_{0}$, $\delta=0.87x_{0}$, with $E_0=\hbar\omega_{0}$ and
$x_{0}=\sqrt{\hbar/\left(m\omega_{0}\right)}$ being the harmonic
oscillator (with frequency $\omega_{0}$) energy and length units.
Consistently, frequency, time and momentum will be measured in units
$\omega_0$, $t_0=1/\omega_0$ and $\hbar k_0 =\sqrt{\hbar m \omega_0}$.

After the trapping potential well is opened, i.e., changed to the
form of equation (\ref{eq:V}), the atoms begin to escape from the
trapping well due to quantum tunnelling. The dynamic is studied by
solving the time dependent nonlinear Schr\"{o}dinger equations (\ref{eq:GPdown})
and (\ref{eq:GPup}) with a Crank-Nicolson scheme.

To characterize the tunnelling escape, we define the number of escaped spin-down/up
atoms as
\begin{equation}
N_{\downarrow,\uparrow}^{out}=\int_{x_{f}}^{\infty}\left|\psi_{\downarrow,\uparrow}\left(x\right)\right|^{2}dx,\label{eq:Nout}
\end{equation}
where $x_{f}=3x_{0}$ is the demarcation point between the potential
area and free space. And the total escaped atoms number is a sum of the
two spin components $N^{out}=N_{\downarrow}^{out}+N_{\uparrow}^{out}$.
It should be noted, the total wave-function is normalized to $1$
(as previously mentioned), therefore the escaped atoms number defined
here is indeed a normalized atoms number or in other words --- fraction
to the total atoms number.

\section{Results\label{sec:results}}

\begin{figure*}
\begin{centering}
\includegraphics{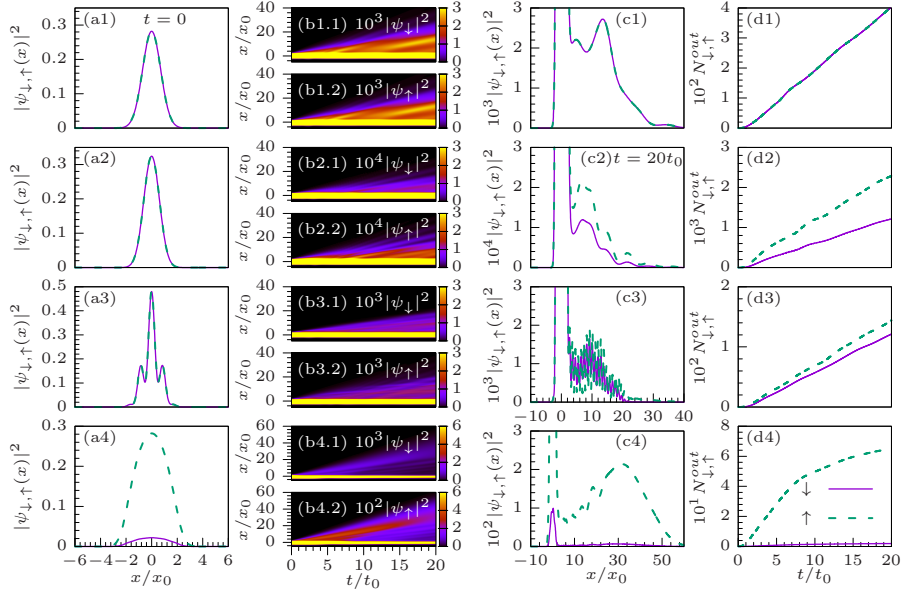}
\par\end{centering}
\caption{Examples of tunnelling escape dynamic for SO coupled BEC in different
phases. Figures (a1-4) are the initial wave packets. Figures (b1.1-4.1),
(b1.2-4.2) are the evolution of spin-down and spin-up atomic densities.
Figures (c1-4) are wave packets at time $t=20 t_0$. Figures (d1-4) are
the escaped spin-down and spin-up atoms number (normalized) during
the evolution. In Figures (a1-d1), there only exists Rabi coupling,
$\Omega=8 \omega_0$, the SO coupling and contact interaction are set to zero,
$p_{c}=0$, $g_{0}=g_{1}=0$. (a2-d2), (a3-d3), (a4-d4) are figures
for initial wave packets in ``single minimum'', ``stripe'' and
``separated'' phases respectively. The corresponding parameters
are (a2-d2): $\Omega=8 \omega_0$, $p_{c}=2 \hbar k_0$, $g_{0}=g_{1}=0$; (a3-d3): $\Omega=8 \omega_0$,
$p_{c}=4 \hbar k_0$, $g_{0}=g_{1}=0$; (a4-d4): $\Omega=8 \omega_0$, $p_{c}=2 \hbar k_0$, $g_{0}=10 E_0 x_0$,
$g_{1}=9 E_0 x_0$. In all these figures, a violet solid line stands for the
spin-down component, while a green dashed line stands for the spin-up
component.
\label{fig:EscapeDynamic}}
\end{figure*}

\begin{figure}
\begin{centering}
\includegraphics{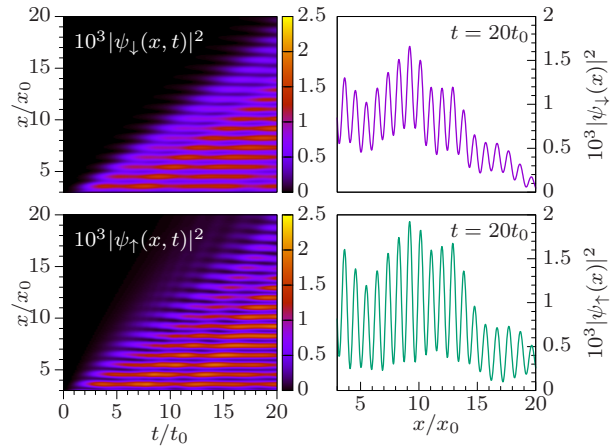}
\par\end{centering}
\caption{Details of figure \ref{fig:EscapeDynamic} (b3.1), (b3.2) and (c3),
pulsive tunnelling escape feature of a stripe phase wave packet. The
left two panels are enlargement of figure \ref{fig:EscapeDynamic}
(b3.1) and (b3.2) in the range of $x\in\left(3x_0,20x_0\right)$. The two
lines in figure \ref{fig:EscapeDynamic} (c3) are enlarged as
the right two panels --- top panel for $\left|\psi_{\downarrow}\left(x,t=20 t_0\right)\right|^{2}$,
while bottom panel for $\left|\psi_{\downarrow}\left(x,t=20 t_0\right)\right|^{2}$.\label{fig:Enlargement}}
\end{figure}

Firstly, for compare purpose and most basic understanding of the system,
we examine the no SO coupling case. Without any SO coupling ($p_{c}=0$),
the symmetry between spin-down and spin-up components suggests the
following form of wave function
\begin{equation}
\Psi\left(x,t\right)=\psi\left(x,t\right)\left(1,-1\right)^{T} e^{i\Omega t}.\label{eq:NoSOCWavefunction}
\end{equation}
Substituting it into nonlinear Schr\"{o}dinger equations (\ref{eq:GPdown})
and (\ref{eq:GPup}), dynamic of the system can be simplified to a
single governing equation
\begin{eqnarray}
i\hbar\frac{\partial}{\partial t}\psi\left(x,t\right) & = & \left[\frac{p_{x}^{2}}{2m}+U\left(x\right)\right]\psi\left(x,t\right)\nonumber \\
 & + & \frac{g_{0}+g_{1}}{2}\left|\psi\left(x,t\right)\right|^{2}\psi\left(x,t\right),\label{eq:NoSOCEquation}
\end{eqnarray}
which has no dependence on the Rabi frequency $\Omega$. This is to
say that without presence of SO coupling, Rabi coupling will not affect
dynamic of the system solely. This fact is also confirmed by our numerical
results. In the first row (a1-d1) of figure \ref{fig:EscapeDynamic},
we set SO coupling parameter $p_{c}=0$, and for each spin components
show the initial wave packet, evolution of atomic density, wave packet
at time $t=20 t_0$ and escaped atoms number during the evolution. We see
that for all these quantities the spin-up and spin-down components
have the same values.

As SO coupling is added to the system, symmetry between spin-down
and spin-up components is broken, see figure \ref{fig:EscapeDynamic}
(a2-d4). For single minimum (figures a2-d2) and stripe phases (figures
a3-d3), although the two spin components have the same initial wave
packets, they gain an evident difference during the evolution. And
for separated phase (figures a4-d4), both the initial wave packets
and their subsequent evolutions for the two spin components are very
different. Interaction is essential in achieving the separated phase
\cite{Li2012}, and the repulsive interaction can considerably reinforce
the escape of the condensate \cite{Huhtamaki2007,Potnis2017}, resulting
a much larger value of escaped atoms number and atomic density compared
to the non-interacting cases.

In figure \ref{fig:EscapeDynamic}, we also noticed that an initial
single minimum or separated phase wave packet escapes from the trapping
well continuously; while because of interference between $+p_{0}$ and
$-p_{0}$ momentum components, an initial stripe phase wave packet
escapes from the trapping well as a series of pulses (for a better
sight, see figure \ref{fig:Enlargement} in which (b3.1), (b3.2),
(c3) of figure \ref{fig:EscapeDynamic} are enlarged to shown details
of the pulsive property). This indicates tunneling escape of a SO
coupled BEC from trapping well may be useful in realizing an atom
laser which can be operated in both continuous and pulsed modes.

\begin{figure}
\begin{centering}
\includegraphics{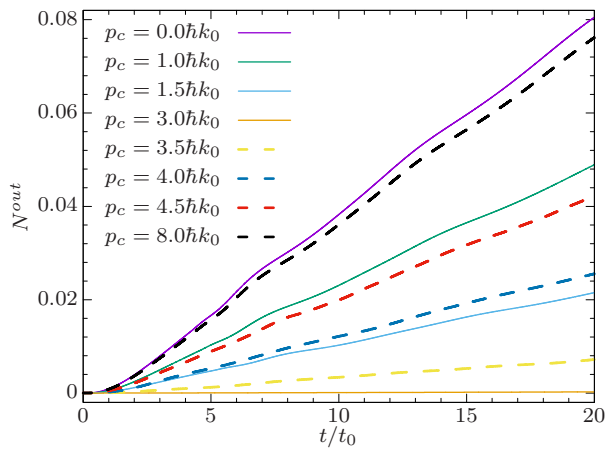}
\par\end{centering}
\caption{Escaped atoms number for different SO coupling strength in the non-interacting
case. The solid lines represent $p_{c}=0.0,1.0,1.5,3.0 \hbar k_0$ (single minimum
phase), while the dashed lines represent $p_{c}=3.5,4.0,4.5,8.0 \hbar k_0$
(stripe phase), respectively. For all the lines, Rabi coupling and
interaction strength are $\Omega=8 \omega_0$, $g_{0}=g_{1}=0$. 
\label{fig:NoutNonInteracting}}
\end{figure}

\begin{figure}
\begin{centering}
\includegraphics{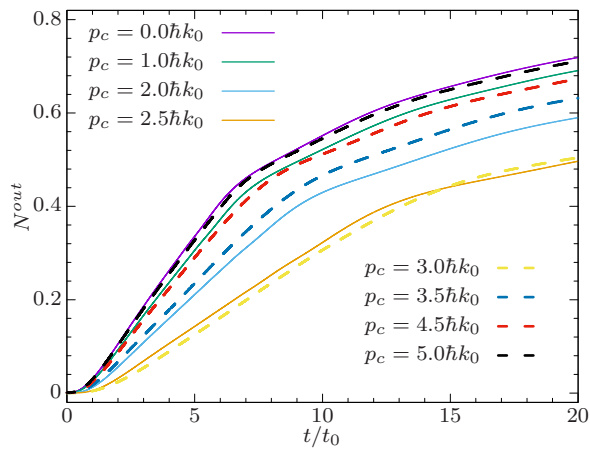}
\par\end{centering}
\caption{Escaped atoms number for different SO coupling strength in the interacting
case. The solid lines represent $p_{c}=0.0,1.0,2.0,2.5 \hbar k_0$ (single minimum
phase), the dashed lines represent $p_{c}=3.0,3.5,4.5,5.0 \hbar k_0$ (separated
phase). The Rabi coupling and interaction strength are $\Omega=8 \omega_0$
and $g_{0}=10E_0 x_0$, $g_{1}=9E_0 x_0$ for all the lines. 
\label{fig:NoutInteracting}}
\end{figure}

From figure \ref{fig:EscapeDynamic}, one can also somewhat see
that SO coupling will affects the escaping speed of the condensate.
This fact can be more clearly demonstrated by figure \ref{fig:NoutNonInteracting}
where we plot the escaped atoms number $N^{out}$ as a function of
time for different SO coupling strength in the case of no inter-atom
interaction. From the figure, we see the line for $p_{c}=0$ (no
SO coupling) locates above all the other ones, i.e., without SO coupling
there are the most escaped atoms, that is to say SO coupling has
a suppressing effect on tunnelling escape of atoms. In the figure, we also
noticed that in the single minimum phase ($p_{c}=0.0,1.0,1.5,3.0 \hbar k_0$), a stronger
SO coupling tends to more reduce the tunnelling escaped atoms number;
while in the stripe phase ($p_{c}=3.5,4.0,4.5,8.0 \hbar k_0$), thing is quite the
opposite --- as the coupling gets stronger, it tends to less reduce
the tunneling escaped atoms number. Around the phase transition point
$p_{c}=3 \hbar k_0$, the escaped atoms number reaches a minimum with value almost
zero. Using this feature, an efficient matter wave switch may be realized.
In the case with inter-atom interaction, a similar suppression phenomenon
can be observed as well (see figure \ref{fig:NoutInteracting}). In
single minimum phase ($p_{c}=0.0,1.0,2.0,2.5\hbar k_0$), the escaped atoms number
also decreases with the increasing of coupling strength. As the SO
coupling getting stronger ($p_{c}=3.0,3.5,4.5,5.0\hbar k_0$), the system goes
into separated phase. In separated phase, the escaped atoms number
increases with the increasing of coupling strength. However, due to
the escape reinforcement caused by repulsive interaction, even near
the phase transition point a sufficient large amount of atoms will
escape the trapping well.

In strong SO coupling limit, according to the dynamical equations
(\ref{eq:GPdown}) and (\ref{eq:GPup}), spin-down/up component tend
to carry a large gauge momentum in the negative/positive direction.
Thus, we can assume the wave function approximately having the following
form
\begin{eqnarray}
\Psi\left(x,t\right) & = & \frac{1}{\sqrt{2}}\phi\left(x,t\right)\left(1,0\right)^{T} e^{-ip_{c}x/\hbar}\nonumber \\
 &  & -\frac{1}{\sqrt{2}}\phi\left(x,t\right)\left(0,1\right)^{T} e^{ip_{c}x/\hbar},\label{eq:LimitWavefunction_mixed}
\end{eqnarray}
for stripe phase, and
\begin{equation}
\Psi\left(x,t\right)=\phi\left(x,t\right)\left(1,0\right)^{T} e^{-ip_{c}x/\hbar},\label{eq:LimitWavefunction_seperated}
\end{equation}
for separated phase, with $\phi\left(x,t\right)$ being a spatially
slow varying wave function. Inserting them into nonlinear coupled
Schr\"{o}dinger equations (\ref{eq:GPdown}) and (\ref{eq:GPup}),
we found that for the stripe phase wave packet, $\phi\left(x,t\right)$
obeys the same equation as non-coupling equation (\ref{eq:NoSOCEquation});
while for the separated phase wave packet, $\phi\left(x,t\right)$
obey a slightly different equation
\begin{eqnarray}
i\hbar\frac{\partial}{\partial t}\phi\left(x,t\right) & = & \left[\frac{p_{x}^{2}}{2m}+U\left(x\right)\right]\phi\left(x,t\right)\nonumber \\
 &  & +g_{0}\left|\phi\left(x,t\right)\right|^{2}\phi\left(x,t\right).\label{eq:LimitSeperatedEquation}
\end{eqnarray}
When inter and intra-components interaction strength are close to
each other $g_{0}\approx g_{1}$, equation (\ref{eq:LimitSeperatedEquation})
is approximately the same as non-coupling equation (\ref{eq:NoSOCEquation}).
So, for strong SO coupling, the tunnelling escape dynamic will go
back to the non-coupling case. This conclusion is also demonstrated
by the fact that in figure \ref{fig:NoutNonInteracting} lines for
$p_{c}=0,8\hbar k_0$ and in figure \ref{fig:NoutInteracting} lines for $p_{c}=0,5\hbar k_0$
are nearly overlapping with each other.

\begin{figure}
\begin{centering}
\includegraphics{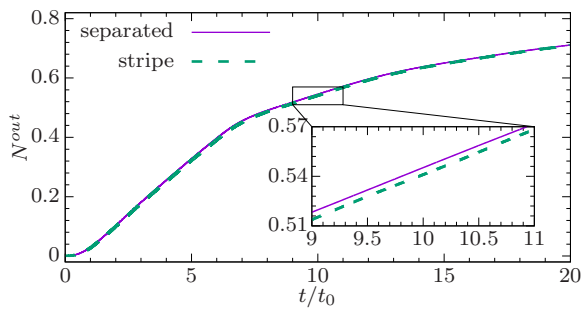}
\par\end{centering}
\caption{Comparison of escaped atoms number for separated (violet solid) and
stripe (green dashed) wave packet. For both lines, the parameters
are $p_{c}=5 \hbar k_0$, $\Omega=8 \omega_0$, $g_{0}=10 E_0 x_0$, $g_{1}=9 E_0 x_0$. Under such parameters,
the initial ground state is in separated phase with energy $E=-12.03 E_0$.
The initial stripe phase state is a nearly degenerate state with a
slightly higher energy $E=-11.91 E_0$. 
\label{fig:SeparatedAndStripe}}
\end{figure}

For inter/intra-component interaction parameters $g_{0}>g_{1}$, as
repulsive interaction will reinforce the tunnelling escape, equations
(\ref{eq:NoSOCEquation}) and (\ref{eq:LimitSeperatedEquation}) suggest
that more atoms will escape the trapping well for a wave packet in
separated phase than in stripe phase. This can be seen from figure
\ref{fig:SeparatedAndStripe}, where we compared the escaped atoms
number during the evolution for initial separated (violet solid line)
and stripe (green dashed line) phase wave packets under parameters
$p_{c}=5\hbar k_0$, $\Omega=8\omega_0$, $g_{0}=10 E_0 x_0$, $g_{1}=9E_0 x_0$. Here, it should
be pointed out that under these parameters, the ground state really
fall in the separated phase with energy $E=-12.03 E_0$. But since the
system is nearly degenerate, a stripe phase wave packet can also been obtained
with a slightly higher energy $E=-11.91 E_0$.

\section{Summary\label{sec:summary}}

In summary, we have analyzed the tunnelling escape of a SO coupled
BEC wave packet from a trapping well. Our results show that the dynamics
of the system are quite different in different phases: a single minimum
or separated phase wave packet escape the trapping well continuously,
while a stripe phase one escape in a pulsed manner. This feature may
be used in realizing atom laser which can be operated both in continuous
and pulsed modes. We also found that SO coupling can suppress the
tunnelling escape of atoms. Especially, in non-interacting case, by
tuning SO coupling strength to the transition point between single
minimum and stripe phase, the tunnelling escape can be almost totally
suppressed, thus a matter wave switch may potentially be realized.
In the strong SO coupling limit, the dynamics go back to the non-coupling
case except for a phase factor.

\section*{Acknowledgment}

The author thanks Prof. Dong Guangjiong from East China Normal University
for very helpful conversations. This work is supported by National Natural Science Foundation of China
(Grant No. 11847059 and 11874127).

\end{document}